\documentclass[11pt]{report}

\usepackage{amsmath,amsthm,amssymb}
\usepackage{graphicx}
\usepackage{subcaption}
\usepackage{listings}

\newtheorem{thm}{Theorem}[section]

\newtheorem{rem}[thm]{Remark}

\newtheorem{algo}[thm]{Algorithm}
\numberwithin{equation}{section}

\newcommand{\R}{\mathbb{R}}

\title{Kalman Filter, Unscented Filter and Particle Flow Filter on Non-linear Models}
\author{
	Author: Yan Zhao\\
	Advisor: prof. Zhongqiang Zhang\\
}
\date{}

\begin{document}
\maketitle

\tableofcontents
\newpage

\begin{abstract}
Filters, especially wide range of Kalman Filters have shown their impacts on predicting variables of stochastic models with higher accuracy then traditional statistic methods. Updating mean and covariance each time makes Bayesian inferences more meaningful. In this paper, we mainly focused on the derivation and implementation of three powerful filters: Kalman Filter, Unscented Kalman Filter and Particle Flow Filter. Comparison for these different type of filters could make us more clear about the suitable applications for different circumstances.     
\end{abstract}

\chapter{ Kalman Filter}
Kalman Filter, also called the Linear Quadratic Estimator(LQE), has been used to minimize the estimation error for unknown variables in  noisy stochastic system. Kalman Filter works recursively to update estimation  by inputting observed measurements over time. It contains two models, the first is Observation model and the second is Measurement model. Observation model, involving Plant noise, has been used to generate prior estimation for current state variables; Measurement model, including observation noise, has been used to update the estimation and generate posterior estimation. Kalman Filter has wide applications, such as predicting natural weather and prices of traded commodities. It also has been used to monitor complex dynamic systems, like signal processing in GPS and motion monitoring in robotics. Kalman Filter works perfectly in linear model, and the extended versions of extended Kalman Filter and Unscented Kalman Filter have been applied to non-linear problems.

\subsection{Linear Dynamic Systems in Discrete Time}
We suppose that the stochastic systems can be presented by the following: \\
Plant model: 
\begin{equation}
x_{k} = \phi_{k-1} x_{k-1} + w_{k-1} \quad with \ w_{k}  \backsim N(0,Q_{k}) \tag{1.1}
\end{equation}
Measurement model:
\begin{equation}
z_{k} = H_{k}x_{k} + v_{k} \quad with \ v_{k} \backsim N(0,R_{k}) \tag{1.2}
\end{equation}
$v_{k}$ and $w_{k}$ are assumed as independent normal random processes with mean of zero. $x_{k}$ has known initial value of $x_{0}$ and known initial covariance matrix $P_{0}$. The goal is to find the estimations of $\hat x_{k}$ presented by function of $z_{k}$ such that the mean-squared error is minimized. Denote $P_{k(-)}$ as the prior covariance matrix for x at time k, $P_{k(+)}$ as the posterior covariance matrix for x at time k, $\bar K_k$ as Kalman gain at time k, $\hat x_{k(-)}$ as the prior estimate of $x_k$ and $\hat x_{k(+)}$ as the posterior estimate of $x_k$. By using orthogonality, we can prove the following updating equations:
\begin{equation}
P_{k(-)} = \phi_{k-1} P_{(k-1)(+)} \phi_{k-1}^T + Q_{k-1} \tag{1.3} 
\end{equation}
\begin{equation}
\bar K_k = P_{k(-)} H_{k}^T [H_k P_{k(-)} H_k^T + R_{k}]^{-1} \tag{1.4}
\end{equation}
\begin{equation}
 P_{k(+)} = [I - \bar K H_k] P_{k(-)} \tag{1.5} 
\end{equation}
\begin{equation}
\hat x_{k(-)} = \phi_{k-1} \hat{x}_{(k-1)(+)} \tag{1.6} 
\end{equation}
\begin{equation}
\hat x_{k(+)} = \hat x_{k(-)} + \bar K [z_k - H_k \hat x_{k(-)}] \tag{1.7}
\end{equation}

\subsection{Example of Application}
Consider a dividend yield and S\&P real return model for stocks, in which $X_n$ is dividend yield, $\delta R_n$ is real return and $Y_n$ is a two-dimensional vector for the observation of $X_n$ and $\delta R_n$ from year 1945 to 2010. $\Delta W_{1,n}, \Delta W_{2,n}$ are independent Brownian motion increments with 
$$\Delta W_{i,n} = W_{i,n+1} - W_{i, n}, i =1,2$$ 
$B_{1,n}, B_{2,n}$ are also independent Brownian motion increments. k, $\theta$, $\sigma$, $\mu$, a, $\rho$, $Q_1$ and $ Q_2$ are parameters with the given values as following:

\begin{table}[h!]
  \begin{center}
    \caption{Parameters}
    \label{tab:table1}
    \begin{tabular}{cccccccc}\\
	\hline
      k & $\theta$ & $\sigma$ & $\mu$ & a & $\rho$ & $Q_1$ & $Q_2$ \\
      \hline
      2.0714 & 2.0451 & 0.3003 & 0.1907 & 0.9197 & 1.6309 & 0.0310 & -0.8857\\
	\hline
    \end{tabular}
  \end{center}
\end{table}

$$ Z_n = \begin{pmatrix} X_n \\ \delta R_n \end{pmatrix} = \begin{pmatrix} \frac{1}{1+k}X_{n-1}+\frac{k\theta}{1+k}+\frac{\sigma}{1+k}\sqrt{X_{n-1}} \Delta W_{1,n} \\ \mu X_n+a\sqrt{X_{n-1}}\left (\rho \Delta W_{1,n} +\sqrt{1-\rho^2} \Delta W_{2,n} \right) \end{pmatrix} $$

$$Y_n = \begin{pmatrix} Y_{1,n} \\ Y_{2,n} \end{pmatrix}  = \begin{pmatrix}  X_n +Q_1B_{1,n} \\ \delta R_n +Q_2B_{2,n}  \end{pmatrix}$$
\\
Rewriting $Z_n$ and $Y_n$ are necessary, as the Observation and Measurement model showing that, $Z_n$ is the function of $Z_{n-1}$ and $Y_n$ is the function of $Z_n$. \\
First, let's rewrite $Z_n$. We can see that $X_{n}$ is represented by $X_{n-1}$, which is the element of vector $Z_{n-1}$, while $\delta R_{n}$ is represented by $X_{n}$. So we need to rewrite $\delta R_n$ as the term of $X_{n-1}$:
\begin{equation}
\delta R_n = \frac{\mu}{1+k} X_{n-1} + \frac{\mu k \theta}{1+k} + (\frac{\mu \sigma \sqrt{x_{n-1}}}{1+k} + a \rho \sqrt{X_{n-1}}) \Delta W_{1,n} + a \sqrt{X_{n-1}} \sqrt{1-\rho^2} \Delta W_{2,n} \tag{1.8}
\end{equation}
Then,
\begin{equation}
Z_n = \begin{pmatrix} X_n\\ \delta R_n  \end{pmatrix} = \begin{pmatrix} \frac{1}{1+k} \ 0 \\ \frac{\mu}{1+k} \ 0 \end{pmatrix} \begin{pmatrix} X_{n-1} \\ \delta R_{n-1} \end{pmatrix} + \begin{pmatrix} \frac{k\theta}{1+k} \\  \frac{\mu k \theta}{1+k} \end{pmatrix} + \sqrt{X_{n-1}} \begin{pmatrix}\frac{\sigma}{1+k} \ \ \ \ \ \ 0 \\ \frac{\mu \sigma}{1+k} +a\rho \ \ \ \ a \sqrt{1-\rho^2} \end{pmatrix}  \begin{pmatrix} \Delta W_{1,n} \\ \Delta W_{2,n} \end{pmatrix} \tag{1.9}
\end{equation}
Denote $$\Phi = \begin{pmatrix} \frac{1}{1+k} \ 0 \\ \frac{\mu}{1+k} \ 0 \end{pmatrix} $$  $$ D = \begin{pmatrix} \frac{k\theta}{1+k} \\ \frac{\mu k \theta}{1+k} \end{pmatrix}$$  $$C = \begin{pmatrix} \frac{\sigma}{1+k} \ \ \ \ \ \ 0 \\ \frac{\mu\sigma}{1+k} + a\rho \ \ \ a\sqrt{1-rho^2} \end{pmatrix} $$  $$ W_n = \begin{pmatrix} \Delta W_{1,n} \\ \Delta W_{2,n} \end{pmatrix}$$
As a result, we can write $Z_n$ as:
\begin{equation}
Z_n = \Phi_{n-1} Z_{n-1} + D + \sqrt{X_{n-1}}CW_n \tag{1.10}
\end{equation}
Next is to rewrite $Y_n$:\\
Denote $$H_n = \begin{pmatrix} 1 \ \ 0 \\ 0 \ \ 1 \end{pmatrix} \ $$ $$V = \begin{pmatrix} Q_1 \ \ 0 \\ 0 \ \ Q_2 \end{pmatrix}$$ $$B_n = \begin{pmatrix} B_{1,n} \\ B_{2,n} \end{pmatrix}$$
We can rewrite $Y_n$ as :
\begin{equation}
Y_n = H_nZ_n + VB_n \tag{1.11}
\end{equation}

\subsection{Solving for Kalman Gain}
The optimal updated estimate $\hat Z_{n(+)}$ is a linear function of a priori estimate $\hat Z_{n(-)}$ and measurement $Y_k$, that is, 
\begin{equation}
\hat Z_{n(+)} = K^1_n \hat Z_{n(-)} + \bar K_n Y_n \tag{1.12}
\end{equation}
$K^1_n$ and $\bar K_n$ are unknown yet. We seek values of $K^1_n$ and $\bar K_n$ such that the estimate $\hat Z_{n(+)}$ satisfies the orthogonality principle:
\begin{equation}
E\langle [Z_n - \hat Z_{n(+)}] Y_i^T \rangle = 0,\ for \ i = 1,2,...n-1 \tag{1.13}
\end{equation} 
If one expand $Z_n$ from equation(1.1) and $Z_{n(+)}$ from equation(1.12) into equation(1.13), then one will obverse:
\begin{equation}
E \langle [\Phi_{n-1} Z_{n-1} + D + \sqrt{X_{n-1}}CW_n - K_n^1 \hat Z_{n(-)}- \bar K_n Y_n] Y_i^T \rangle = 0,\ for \ i = 1,2, ...n-1 \tag{1.14}
\end{equation}
Since $W_n$ and $V_n$ are uncorrelated, it follows that $E \langle W_nY_i^T\rangle  = 0 \ for \ 1 \leq i \leq n-1$. Using this result, one can get obtain the following result:
\begin{equation}
E \langle [\Phi_{n-1} Z_{n-1} + D - K_n^1 \hat Z_{n(-)}- \bar K_n Y_n] Y_i^T \rangle = 0,\ for \ i = 1,2, ...n-1 \tag{1.15}
\end{equation}
Then by substituting $Y_n$ using equation (1.11), one can get
\begin{equation}
E \langle [\Phi_{n-1} Z_{n-1} + D - K_n^1 \hat Z_{n(-)}- \bar K_n  H_nZ_n - \bar K_n VB_n] Y_i^T \rangle = 0,\ for \ i = 1,2, ...n-1 \tag{1.16}
\end{equation}
Then equation(1.16) can be changed to the form
\begin{equation}
\begin{aligned}
& \Phi_{n-1}E \langle Z_{n-1} Y_i^T \rangle + D E\langle  Y_i^T \rangle - K_n^1 E \langle \hat Z_{n(-)} Y_i^T \rangle - \bar K_n  H_n E \langle Z_n  Y_i^T \rangle \\
&- \bar K_n V E \langle B_n Y_i^T \rangle = 0, \ for \ i = 1,2, ...n-1  
\end{aligned} \tag{1.17}
\end{equation}
We also know that 
$$E \langle B_n Y_i^T \rangle = 0, \ for \ i = 1,2, ...n-1 $$
Equation (1.17) can be reduced to the form
\begin{equation}
\begin{aligned}
& \Phi_{n-1}E \langle Z_{n-1} Y_i^T \rangle + D E\langle  Y_i^T \rangle - K_n^1 E \langle \hat Z_{n(-)} Y_i^T \rangle - \bar K_n  H_n E \langle Z_n  Y_i^T \rangle = 0, \\
&E \langle [Z_{n}- K_n^1 Z_{n}- \bar K_n  H_nZ_n] Y_i^T \rangle- K_n^1 E\langle [\hat Z_{n(-)} - Z_n]Y_i^T \rangle = 0, \\
&E \langle [Z_{n}- K_n^1 Z_{n}- \bar K_n  H_nZ_n] Y_i^T \rangle =0, \\
&E \langle [I - K_n^1 - \bar K_n  H_n]\rangle E\langle Z_n Y_i^T \rangle =0
\end{aligned} \tag{1.18}
\end{equation}
Equation (1.18) can be satisfied for any given $Z_n$ if 
\begin{equation}
 K_n^1 = I - \bar K_n  H_n, \tag{1.19}
\end{equation}
Thus, $K_n^1$ in equation (1.12) satisfied equation (1.19).\\
Define estimation errors after and before updates
\begin{equation}
\tilde Z_{n(+)} \triangleq \hat Z_{n(+)} -Z_n \tag{1.20}
\end{equation}
\begin{equation}
\tilde Z_{n(-)} \triangleq \hat Z_{n(-)} -Z_n \tag{1.21}
\end{equation}
\begin{equation}
\begin{aligned}
&\tilde Y_{n} \triangleq \hat Y_{n(-)} -Y_n \\
& = H_nZ_{n(-)} + Y_n
\end{aligned} \tag{1.22}
\end{equation}
Since $\tilde Y_{n(-)}$ depends linearly on $Y_n$, from equation (1.13),
\begin{equation}
E\langle [Z_n - \hat Z_{n(+)}] \tilde Y_n^T \rangle = 0 \tag{1.23}
\end{equation}
Substitute $Z_n$, $\hat Z_{n(+)}$, and $\tilde Y_n$ from equations (1.10), (1.12), (1.22)  respectively. Then
$$E \langle \Phi_{n-1} Z_{n-1} + D + \sqrt{X_{n-1}} CW_n -K_n^1 \hat Z_{n(-)} -\bar K_n Y_n] [H_n \hat Z_{n(-)} - Y_n]^T \rangle = 0. $$
By the orthogonality of 
$$E \langle W_n Y_n^T \rangle = E \langle W_n X_{n(-)}^T \rangle = 0, $$
We will obtain
$$E \langle \Phi_{n-1} Z_{n-1} + D -K_n^1 \hat Z_{n(-)} -\bar K_n Y_n] [H_n \hat Z_{n(-)} - Y_n]^T \rangle = 0. $$
Substituting for $K_n^1$, $Y_n$ and using equation (1.21)
$$E \langle \Phi_{n-1} Z_{n-1} + D - \hat Z_{n(-)} +\bar K H_n \hat Z_{n(-)} -\bar K_n Z_n - \bar K_n V B_n] [H_n \hat Z_{n(-)} - H_n Z_n - VB_n]^T \rangle = 0, $$
$$E \langle [(Z_n - \hat Z_{n(-)}) - \bar K_n H_n (Z_n - \hat Z_{n(-)}) - \bar K_n V B_n] [H_n (\hat Z_{n(-)} - Z_n) - VB_n]^T \rangle= 0, $$
$$E \langle [-\tilde Z_{n(-)} +\bar K_n H_n \tilde Z_{n(-)} - \bar K_n V B_n] [H_n \tilde Z_{n(-)} - VB_n]^T \rangle= 0, $$
\begin{equation}
\begin{aligned}
& (-I + \bar K_n H_n) E\langle \tilde Z_{n(-)} \tilde Z_{n(-)}^T \rangle H_n^T - (-I + \bar K_n H_n) E\langle \tilde Z_{n(-)}B_n^T\rangle V^T  \\
& - \bar K_n V E\langle B \tilde Z_{n(-)}^T \rangle H_n^T + \bar K_n VE \langle B_n B_n^T \rangle V^T  = 0
\end{aligned} \tag{1.24}
\end{equation}
Using the fact that $E \langle \tilde Z_{n(-)} B_n^T \rangle = E\langle B_n^T \tilde Z_{n(-)}^T \rangle = 0$, this last result will be as follows:
\begin{equation}
(-I + \bar K_n H_n) E\langle \tilde Z_{n(-)} \tilde Z_{n(-)}^T \rangle H_n^T + \bar K_n VE \langle B_n B_n^T \rangle V^T  = 0 \tag{1.25}
\end{equation}
For the second term of equation(1.25) $\bar K_n VE \langle B_n B_n^T \rangle V^T$:
\begin{equation}
\bar K_n \begin{pmatrix} Q_1 \ 0 \\ 0 \ Q_2 \end{pmatrix} E \begin{pmatrix} B_{1n}^2 \ \ B_{2n}B_{1n} \\ B_{1n}B_{2n} \ \ B_{2n}^2 \end{pmatrix} \begin{pmatrix} Q_1 \ 0 \\ 0 \ Q_2 \end{pmatrix} = \bar K_n \begin{pmatrix} Q_1 \ 0 \\ 0 \ Q_2 \end{pmatrix} \begin{pmatrix} 1 \ \ 0 \\ 0 \ \ 1 \end{pmatrix} \begin{pmatrix} Q_1 \ 0 \\ 0 \ Q_2 \end{pmatrix} = \bar K_n V^2 \tag{1.26}
\end{equation}
Plugging the value of equation (1.26) to (1.25):
$$ (-I + \bar K_n H_n) E\langle \tilde Z_{n(-)} \tilde Z_{n(-)}^T \rangle H_n^T + \bar K_n V^2 = 0, $$
By definition, the error covariance matrix is $P_{n(-)} = E\langle \tilde Z_{n(-)} \tilde Z_{n(-)}^T \rangle$, it satisfies the equation:
$$  (-I + \bar K_n H_n)P_{n(-)} H_n^T +  \bar K_n V^2 = 0, $$ 
$$ \bar K_n (H_n P_{n(-)} H_n^T + V^2) = P_{n(-)}H_n^T, $$
And therefore, Kalman gain can be expressed as:
\begin{equation}
\bar K_n =  P_{n(-)}H_n^T (H_n P_{n(-)} H_n^T + V^2) ^{-1}, \tag{1.27}
\end{equation}
which is the solution we want to seek as a function of priori covariance before update.

\subsection{Solving for Priori and Posterior Estimation}
By definition, the priori estimation
\begin{equation}
\hat Z_{n(-)} = \Phi_{n-1} \hat Z_{n(+)} + D. \tag{1.28}
\end{equation}
By substituting equation (1.19) into equation (1.12), one obtains the equations
$$ \hat Z_{n(+)} = (I - \bar K_n H_n) \hat Z_{n(-)} + \bar K_n Y_n , $$
\begin{equation}
\hat Z_{n(+)} = \hat Z_{n(-)} + \bar K_n (-H_n \hat Z_{n(-)} + Y_n) \tag{1.29}
\end{equation}
Therefore, the posterior estimation we want to seek is a function of priori estimation and kalman gain.

\subsection{Solving for Prior and Posterior Covariance}
One can derive a formula for posterior covariance, which is 
\begin{equation}
P_{n(+)} = E \langle \tilde Z_{n(+)}\tilde Z_{n(+)}^T \rangle \tag{1.30}
\end{equation}
By plugging equation (1.29) to equation (1.20), one obtains the equations
\begin{eqnarray}
\tilde Z_{n(+)} &= &\hat Z_{n(+)} - Z_n =  \hat Z_{n(-)} - \bar K_n H_n \hat Z_{n(-)} + \bar K_n Y_n -Z_n \notag  \\
& =& \hat Z_{n(-)} - \bar K_n H_n \hat Z_{n(-)} + \bar K_n H_n Z_n - \bar K_n V B_n -Z_n  \notag \\
& =& (\hat Z_{n(-)} Z_n) - \bar K_n H_n (\hat Z_{n(-)} - Z_n) + \bar K_n V B_n   \notag \\ 
& =& (I - \bar K_n H_n) \tilde Z_{n(-)} + \bar K_n V B_n
  \label{1.31}
\end{eqnarray}
By substituting equation (1.31) into equation (1.30) and noting that $E \langle \tilde Z_{n(-)} B_n^T \rangle = 0$, one obtains
\begin{equation}
\begin{aligned}
&P_{n(+)} = E \langle [(I - \bar K_n H_n) \tilde Z_{n(-)}+\bar K_n V B_n] [(I - \bar K_n H_n) \tilde Z_{n(-)}+\bar K_n V B_n]^T \rangle \\
&= E \langle (I - \bar K_n H_n) \tilde Z_{n(-)} \tilde Z_{n(-)}^T (I - \bar K_n H_n)^T + \bar K_n V B_n B_n^T V^T \bar K_n^T \rangle \\
&=(I - \bar K_n H_n) P_{n(-)}  (I - \bar K_n H_n)^T + \bar K_n V^2 \bar K_n^T \\
&= P_{n(-)} - \bar K_n H_n P_{n(-)} - P_{n(-)} H_n^T \bar K^T + \bar K_n H_n P_{n(-)} H_n^T \bar K_n^T + \bar K_n V^2 \bar K_n^T \\
&= (I - \bar K_n H_n) P_{n(-)} - P_{n(-)} H_n^T \bar K_n^T + \bar K_n (H_n P_{n(-)} H_n^T + v^2) \bar K^T \\
&= (I - \bar K_n H_n) P_{n(-)} - P_{n(-)} H_n^T \bar K_n^T +  P_{n(-)} H_n^T \bar K_n^T \\
&= (I - \bar K_n H_n) P_{n(-)}
\end{aligned} \tag{1.32}
\end{equation}
This is the final form of posterior covariance, which shows the effects of kalman gains on priori covariance. 
Respectively, the definition of prior covariance 
\begin{equation}
P_{n(-)} = E \langle \tilde Z_{n(-)}\tilde Z_{n(-)}^T \rangle \tag{1.33}
\end{equation}
By plugging equation (1.10) and equation (1.28) to equation (1.21), one obtains the equations
\begin{equation}
\begin{aligned}
&\tilde Z_{n(-)} = \Phi_{n-1} \hat Z_{n-1(+)} + D - Z_n \\
&= \Phi_{n-1} \hat Z_{n-1(+)} + D - Z_n - \Phi_{n-1} Z_{n-1} - D - \sqrt{X_{n-1}}CW_n  \\
&= \Phi_{n-1} \tilde Z_{n-1(+)} - \sqrt{X_{n-1}} CW_n
\end{aligned} \tag{1.34}
\end{equation}
Uses the fact that $ E\langle \tilde Z_{n-1} W_{n-1}^T$ to obtain the results
\begin{equation}
\begin{aligned}
&P_{n(-)} = E \langle [\Phi_{n-1} \tilde Z_{n-1(+)} - \sqrt{X{n-1}} CW_n] [\Phi_{n-1} \tilde Z_{n-1(+)} - \sqrt{X{n-1}} CW_n]^T \rangle \\
& = \Phi_{n-1} E \langle \tilde Z_{n-1(+)} \tilde Z_{n-1(+)}^T \rangle \Phi_{n-1}^T + \sqrt{X_{n-1}} C E \langle W_n W_n^T \rangle C^T \sqrt{X_{n-1}} \\
&= \Phi_{n-1} P_{n-1(+)} \Phi_{n-1}^T + X_{n-1} CC^T
\end{aligned} \tag{1.35}
\end{equation}
which gives a priori value of the covariance matrix as a function of the previous posterior covariance. \\
Thus, the update equations for our yield and real return model are listed following:
\begin{equation}
P_{n(-)} = \Phi_{n-1} P_{n-1(+)} \Phi_{n-1}^T + X_{n-1} CC^T \tag{1.35}
\end{equation}

\begin{equation}
\bar K_n =  P_{n(-)}H_n^T (H_n P_{n(-)} H_n^T + V^2) ^{-1}, \tag{1.27}
\end{equation}

\begin{equation}
P_{n(+)} = (I - \bar K_n H_n) P_{n(-)} \tag{1.32}
\end{equation}

\begin{equation}
\hat Z_{n(-)} = \Phi_{n-1} \hat Z_{n(+)} + D. \tag{1.28}
\end{equation}

\begin{equation}
\hat Z_{n(+)} = \hat Z_{n(-)} + \bar K_n (-H_n \hat Z_{n(-)} + Y_n) \tag{1.29}
\end{equation}
The form of equations of example model are similar to equation (1.3) to (1.7), but the differences are because the example model is not strictly linear and noisy parts from plant model are relying on the previous steps.

\subsection{Results for Yield and Real Return Model}
By plugging value of Yield and Real Return from year 1945 to year 2010 to $Y_n$ and setting the initial priori covariance as zero, one can repeat the algorithms listed above to calculate kalman gain 65 times and correspondingly update post covariance and posterior value of estimation. Set posterior estimation as estimation for yield and real return, and one can plot real value and estimation value on the same plot by using same time discretization. The results are showing following:

\begin{figure}[h!]
  \centering
  \begin{subfigure}[b]{0.7\linewidth}
    \includegraphics[width=\linewidth]{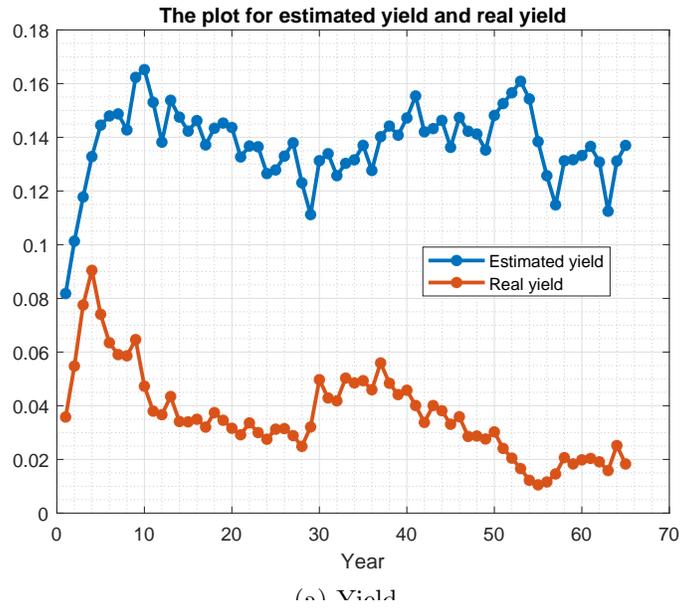}
    \caption{Yield}
  \end{subfigure}
  \begin{subfigure}[b]{0.7\linewidth}
    \includegraphics[width=\linewidth]{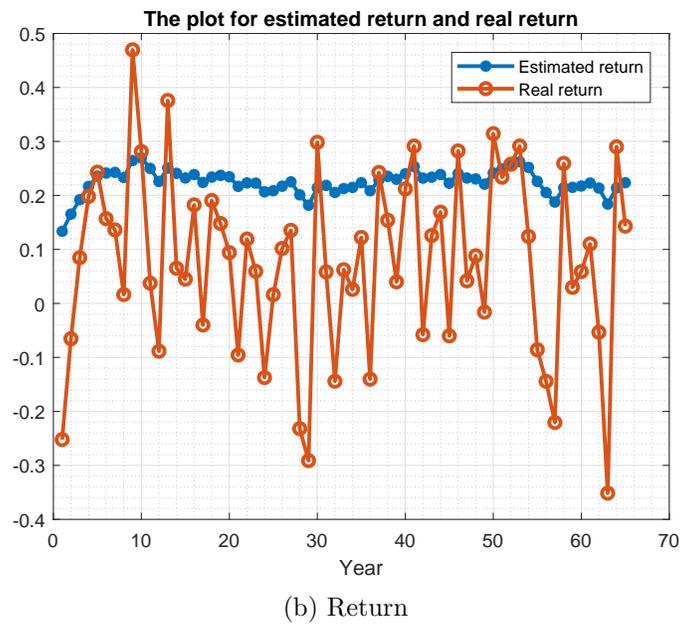}
    \caption{Return}
  \end{subfigure}
  \caption{Kalman Filter Results for Yield and Real Return}
\end{figure}
The results are showing that kalman filter works well in first five to six years with the same trend of move and estimation value approximating to real value. After the fifth year, the value of estimations are far away from real value but keeping the same trend of move. The reason for estimation and real value deviating from fifth year is that the model is non-linear with time. The results confirm that kalman filter perfectly works on linear model and the first several steps of non-linear model, while it works worse on the later part of non-linear model. Thus, the use of extended Kalman filter-- Unscented Kalman filter, is needed to solve this non-linear problem. 

\chapter{Unscented Filtering and Nonlinear Estimation}
The extended Kalman Filter (EKF) has been widely used to deal with non-linear problem. However, it is hard to implement and the results are often inaccurate. the Unscented transformation (UT) has been developed as an improvement to utilize information of mean and covariance to accurate results and make it easier to implement. The method is to select sigma points according to their mean $\mu_x$ and covariance $\sigma_x$ (i.e. choosing data in range of $[-2 \sigma_x, 2\sigma_x]$). The non-linear function is applied to each point to generate a cloud of points. Then transformed mean and covariance can be obtained from calculating mean and variance of those sigma points. There are two advantages of using UT transformation. The first is selected sigma points are no longer randomly chosen but containing information of an unknown distribution, which is sufficient to operate statistic computation. Furthermore, mean and covariance are linearly transformable (i.e. mean $\bar x$ will be $T\bar x$ after operating transformation T, and covariance $\Sigma_x$ will be $T \Sigma_x T^T$) The second is weights for sigma points can be adjusted in ways such that more points around mean can be captured. 

\subsection{General Algorithms for Unscented Kalman Filter}
1) Generating sigma points:\\
Consider a set of sigma points S with given mean and covariance, it contains $(2N_x+1)$ vectors and their associate weights $S = \{i = 0,1,...2N_x: X^{(i)}, W^{(i)} \}$. By convention, $W^{(0)}$ will be the weight on the mean point, which is indexed as the zeroth point
$$X^{(0)} = \bar X$$
$$W^{(0)} = W^{(0)}$$
The other $2N-x$ points lie on the $\sqrt{N_x}$th covariance with half points on the left side of mean and half on the right side of mean
$$X^{(i)} = \bar X + \big (\sqrt{ \frac{N_x}{1-W^{(0)}} \Sigma_x}\big )_i $$
$$ W^{(i)} = \frac{1-W^{(0)}}{2N_x}$$
$$X^{(i+N_x)} = \bar X - \big (\sqrt{ \frac{N_x}{1-W^{(0)}} \Sigma_x}\big )_i $$
$$ W^{(i+N_x)} = \frac{1-W^{(0)}}{2N_x}$$
2) Generating transformed set, which is normally the expectation value through Plant model
$$ \hat X_n^{(i)} = f[X_n^{(i)},\mu_n]. $$
3) Computing predicted mean
$$ \hat \mu_n = \sum_{i=0}^{p} W^{(i)} \hat X_n^{(i)}. $$
4) And computing predicted covariance
$$ \hat K_n = \sum_{i=0}^{p} \{ \hat X_n^{(i)} - \hat \mu_n \}  \{ \hat X_n^{(i)} - \hat \mu_n \}^T. $$
5) Plugging each of the predicted points to observation model
$$ \hat Y_n^{(i)}  = g[X_n^{(i)}]. $$
6) Computing observation mean
$$ \hat Y_n = \sum_{i=0}^{p} W^{(i)} \hat Y_n^{(i)}. $$
7) And computing observation covariance
$$ \hat S_n = \sum_{i=0}^{p} \{ \hat Y_n^{(i)} - \hat Y_n \}  \{ \hat Y_n^{(i)} - \hat Y_n \}^T. $$
8) Finally updating normal Kalman Filter Equations
$$ \mathcal{V}_n = Y_n - \hat Y_n$$
$$ W_n = \hat K_n \hat Y_n^{-1} $$
$$ \mu_n = \hat \mu_n + W_n \mathcal{V}_n $$
$$ K_n = \hat K_n - W_n \hat S_n W_n^T $$

\subsection{Implementation for Yield and Real Return Model}
Here we use the same example of yield and real return model. Expected results are better by implementing Unscented Kalman filter. That is 
\begin{equation}
Z_n = \Phi_{n-1} Z_{n-1} + D + \sqrt{X_{n-1}}CW_n \tag{1.10}
\end{equation}
\begin{equation}
Y_n = H_nZ_n + VB_n \tag{1.11}
\end{equation}
In order to generate sigma points, $W^{(0)} = \frac{1}{3}$ and $2N_x = 400$ have been set. With initial mean of sigma points $\mu_n = H_n^{-1} (Y_n - VB_n)$ and initial covariance as zero matrix, one can repeat the algorithms by using steps listed. Each time by choosing factorization of $\hat K_n$, we can get the covariance of sigma points. When implementing the algorithm, we changed a little bit in step 2. Instead of using expectation, we use the whole function to process sigma particles because the value of noisy parameters are relatively high and it will be better to mimic points adding those noise. Each time we need to guarantee $X_n$ is positive. 

\subsection{Results for Yield and Real Return Model}
Predicted dividend yield matches highly with the real yield from the figure, which means the prediction for yield is pretty sucess. Predicted real return does not match with the real return well but keep the same trend. Reasons for diiference of the results are: First, variance for real return is higher than yield which enlarge the error for mis-allocated sigma points. Sigma points for yield are intensive since it has relatively stable trend with lower variance. Second, for updating each step, real return highly depends on the prediction of yield from previous step, so the predicted error for yield can be exaggerated further.
\begin{figure}[h!]
  \centering
  \begin{subfigure}[b]{0.7\linewidth}
    \includegraphics[width=\linewidth]{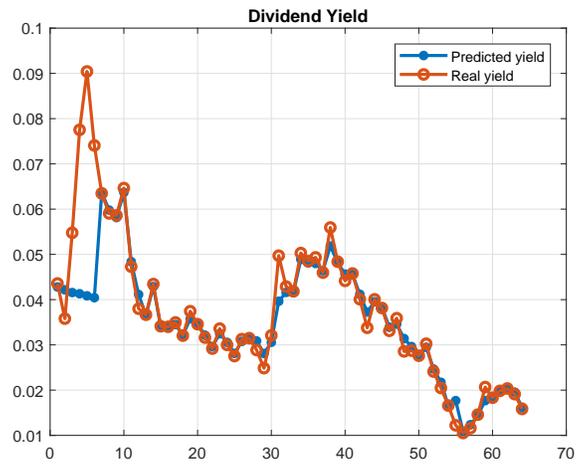}
    \caption{Yield}
  \end{subfigure}
  \begin{subfigure}[b]{0.7\linewidth}
    \includegraphics[width=\linewidth]{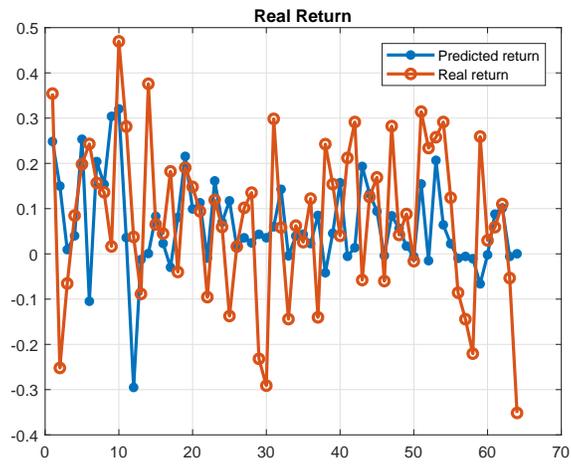}
    \caption{Return}
  \end{subfigure}
  \caption{Unscented Filter Results for Yield and Real Return}
\end{figure}

\chapter{Particle Flow Filter}

Particle Filters have the problem of particle degeneracy caused by Bayesian Rule, especially in dealing with high dimensional state vectors. The algorithm puts particles to wrong places when multiplying prior function with likelihood function. Particle Flow Filter is derived to improve the estimation accuracy in high-dimensional space by involving move functions of particles and it is significantly mitigate the problem of degeneracy. We set each particle in d-dimensional space as a function of $\lambda$ denoting as $x(\lambda)$, in which lambda is continuously changing like time. $\lambda$ starts from 0 and ends up with 1 giving the results of moving from points to next points.

\subsection{Generalized Gromov Method for stochastic Particle Flow Filters}

We start from constructing the stochastic differential equation for flow of particles:
\begin{equation} \label{stochastic flow of particle}
dx = f(x,\lambda)d\lambda + Q(x)^{\frac{1}{2}} dW_{\lambda}
\end{equation} 
Here $f(x, \lambda)$ is the moving function for particles and Q is the covariance matrix of the diffusion $W_\lambda$. $W_\lambda$ is the measurement noise generated according to $\lambda$. \\
In order to get the solution of $f(x, \lambda)$ and Q(x), probability density function $log \ P(x, \lambda)$ is essential to be introduced. We have:
\begin{equation} \label{log probability density function}
\log \ P(x,\lambda) = \log \ g(x) +\lambda \log \ h(x) - \log\ K(\lambda)
\end{equation} 
The generalized probability density function has the form of :
\begin{equation} \label{probability density function}
	p(x, \lambda) = \frac{g(x)h(x)^{\lambda}}{\int_{\mathbb{R}^d} g(x)h(x)^{\lambda}\,dx} =\frac{g(x)h(x)^{\lambda}}{K(\lambda)} , 
\end{equation} 
in which h(x) is the likelihood, g(x) is from part a and $K(\lambda)$ is the norm of product of $g(x) and h(x)^\lambda$. The purpose of $K(\lambda)$ is to normalize the conditional probability density. \\
By using equation \eqref{log probability density function}, one can solve $f(x,\lambda)$ by setting specific $Q(x)$ to simplify the PDE for f. The PDE has the form of :
\begin{equation}
\frac{\partial{\log h}} {\partial x} = - f^T \frac{\partial^2 \log \ P}{\partial x^2} - \frac{\partial div(f)}{\partial x} - \frac{\partial \log P}{\partial x} \frac{\partial f}{\partial x} + \frac{\partial[div(Q\frac{\partial P}{\partial x}) /2P]}{\partial x}
\end{equation}
The simplest way is to set:
\begin{equation} \label{solve for particle flow PDE}
-\frac{\partial div(f)}{\partial x} - \frac{\partial log P}{\partial x} \frac{\partial f}{\partial x} + \frac{\partial[div(Q\frac{\partial P}{\partial x}) /2P]}{\partial x} = 0 
\end{equation} 
Then the solution for $f(x,\lambda)$ is :
\begin{equation} \label{solution for f}
f(x, \lambda) = - [\frac{\partial^2 \log  P(x,\lambda)}{\partial x^2}]^{-1} (\frac{\partial \log h(x)}{\partial x})^T
\end{equation}
According to equation \eqref{solve for particle flow PDE},  the corresponding covariance function Q is:
\begin{equation} \label{solution for Q}
Q = [P-\lambda PH^T(R+\lambda H P H^T)^{-1} HP] H^TR^{-1}H  [P-\lambda PH^T(R+\lambda H P H^T)^{-1} HP]
\end{equation}
where $R$ is the measurement noise covariance matrix, $P$ is the prior covariance matrix, and $H$ is the sensitive matrix in measurement model. \\
In order to keep the solution of $Q$ from equation \eqref{solution for Q} as symmetric matrix, one can implement the following method to symmetry Q immediately:
\begin{equation}
Q = \frac{Q + Q^T}{2}
\end{equation}
 
\begin{algo}(Algorithm for implementing Particle Flow Filter with diffusion)  \upshape

\begin{itemize}
\item a. Use Monte Carlo method randomly choose $N$ particles around observation, and generate particle density function $g(x)$ as prior density function.\\

\item b. Choose suitable $h(x)$ as likelihood function.  

\item c. Compute $p(x,\lambda)$ by Equation \eqref{probability density function}, $p(x, \lambda) = \frac{g(x)h(x)^{\lambda}}{K(\lambda)}$, where 
$K(\lambda)=\int_{\R^d} g(x)h(x)^{\lambda}\,dx$
.  

\item d. Solve the moving function $f(x,\lambda)$ and measurement covariance matrix Q by equation \eqref{solution for f} and \eqref{solution for Q}. That is, 
\begin{equation}
 f(x, \lambda) = - [\frac{\partial^2 \log \ P(x,\lambda)}{\partial x^2}]^{-1} (\frac{\partial \log h(x)}{\partial x})^T.
\end{equation}  
\[Q = [P-\lambda PH^T(R+\lambda H P H^T)^{-1} HP] H^TR^{-1}H  [P-\lambda PH^T(R+\lambda H P H^T)^{-1} HP].\]

\item e. Plug the value of $f(x,\lambda)$ and $Q(x)$, one can derive x by solving the PDE: $ dx = f(x,\lambda)d\lambda + L dW_{\lambda}$, with L= chol(Q). We can use forward Euler scheme 
\begin{equation}   x^{(n+1)} = x^{(n)} + f(x^{(n)}, \lambda_n) \Delta \lambda + L \Delta W_{\lambda}
\end{equation} or implicit Euler scheme
\begin{equation}
 x^{(n+1)} = x^{(n)} + f(x^{(n+1)}, \lambda_{n+1}) \Delta \lambda + L \Delta W_{\lambda}.
\end{equation}  

f. For updating each point, repeat steps from a to e.

\end{itemize}
\end{algo}

 \begin{rem}
 
 Here  $h(x)$ can be any type of distribution but we  consider normal distribution with estimated mean and variance. 
 
\end{rem}
 
 \begin{rem}
 The use of either explicit or implicit Euler method depends on the shape of $f(x,\lambda)$. 
\end{rem}

\subsection{Implementation of Particle Flow Filter}
In our previous dividend yield and S\&P real return model,with the observation model as: 
$$ Z_n = \begin{pmatrix} X_n \\ \delta R_n \end{pmatrix} = \begin{pmatrix} \frac{1}{1+k}X_{n-1}+\frac{k\theta}{1+k}+\frac{\sigma}{1+k}\sqrt{X_{n-1}} \Delta W_{1,n} \\ \mu X_n+a\sqrt{X_{n-1}}\left (\rho \Delta W_{1,n} +\sqrt{1-\rho^2} \Delta W_{2,n} \right) \end{pmatrix} $$
and measurement model as:
$$Y_n = \begin{pmatrix} Y_{1,n} \\ Y_{2,n} \end{pmatrix}  = \begin{pmatrix}  X_n +Q_1B_{1,n} \\ \delta R_n +Q_2B_{2,n}  \end{pmatrix}$$
We can get the particle density function is
$$ g(x_1,x_2) = \frac{1}{2\pi \sigma_1 \sigma_2 \sqrt{1- \rho^2}} e^{-\frac{(x-\mu)^T \Sigma_1 ^{-1} (x-\mu)}{2}} $$
where $\mu$ is sample mean and $\Sigma_1$ is sample covariance 
$$ \Sigma_1 = \begin{pmatrix} \sigma_1^2 \ \ \rho \sigma_1 \sigma_2 \\ \rho \sigma_1 \sigma_2\  \ \sigma_2^2 \end{pmatrix}$$
We can set the likelihood function as
$$ h(x_1,x_2) = \frac{1}{2\pi \sqrt{|\Sigma_2|}} e^{-\frac{(x-m)^T \Sigma_2 ^{-1} (x-m)}{2}} $$
where m is probability mean and $\Sigma_2$ is probability covariance.
Conditional probability density function $P(x, \lambda)$ follows:
$$p(x, \lambda) = \frac{g(x)h(x)^{\lambda}}{|| g(x)h(x)^{\lambda}||} =  \frac{ e^{-\frac{(x-\mu)^T \Sigma_1 ^{-1} (x-\mu) + \lambda (x-m)^T \Sigma_2 ^{-1} (x-m) }{2}}} {K(\lambda)}$$
where $$K(\lambda) =  ||e^{-\frac{(x-\mu)^T \Sigma_1 ^{-1} (x-\mu) + \lambda (x-m)^T \Sigma_2 ^{-1} (x-m) }{2}} ||$$
And then
$$ log P(x,\lambda) = -\frac{(x-\mu)^T \Sigma_1 ^{-1} (x-\mu) + \lambda (x-m)^T \Sigma_2 ^{-1} (x-m) }{2} - log(K(\lambda)) $$
$$ log \ h(x) = log(\frac{1}{2 \pi (\Sigma_2)^{1/2}}) -\frac{(x-m)^T \Sigma_2 ^{-1} (x-m)}{2} $$
$$ \frac{\partial ^2(log P(x,\lambda))} {\partial x^2} = -\Sigma_1 ^{-1} - \lambda \Sigma_2 ^{-1}$$
$$ \frac{\partial (log \ h(x))}{\partial x} = -\Sigma_2^{-1} (x-m)$$
Moving function $f(x, \lambda)$ is
$$ f(x, \lambda) = - [\frac{\partial^2 log \ P(x,\lambda)}{\partial^2 x}]^{-1} (\frac{\partial log \ h(x)}{\partial x}) = -[-\Sigma_1 ^{-1} - \lambda \Sigma_2 ^{-1}]^{-1} [-\Sigma_2^{-1} (x-m)] $$
$$= -[\Sigma_1 ^{-1} + \lambda \Sigma_2 ^{-1}]^{-1} \Sigma_2^{-1} (x-m)$$
According to equation \eqref{solution for Q}, the corresponding $Q(x)$ in this case is:
$$Q = [P-\lambda P(V+\lambda P)^{-1} P] V^{-1} [P-\lambda P(V+\lambda P)^{-1}P], $$
where $P$ is the prior covariance, which has the form from Kalman Filter:
$$P_{n(-)} = \Phi P_{n-1(+)} \Phi^T + (1 \ , \ 0) x^{(n-1)} CC^T,$$
and $L(x) = \sqrt{Q(x)} $. 
Then update x with respect to $\lambda$ by Backward Euler:
$$x^{(n+1)} = x^{(n)} + f(x^{(n+1)}, \lambda_{n+1}) \Delta \lambda + L(x^{(n)}) \Delta W_{\lambda}$$
Subtracting $m$ on each side, one can get
$$x^{(n+1)} - m = (x^{(n)}- m) -[\Sigma_1 ^{-1} + \lambda_{n+1} \Sigma_2 ^{-1}]^{-1} \Sigma_2^{-1} (x^{(n+1)}-m) \Delta \lambda + L(x^{(n)}) \Delta W_{\lambda}.$$
Set $y^{(n+1)} = x^{(n+1)} - m$ and $y^{(n)} = x^{(n)} - m$, the equation becomes
$$ y^{(n+1)} = y^{(n)} - [\Sigma_1 ^{-1} + \lambda \Sigma_2 ^{-1}]^{-1} \Sigma_2^{-1} y^{(n+1)} \Delta \lambda + L(x) \Delta W_{\lambda}.$$
$$(I + \Delta \lambda [\Sigma_1 ^{-1} + \lambda \Sigma_2 ^{-1}]^{-1} \Sigma_2^{-1}) y^{(n+1)} = y^{(n)} + L(x) \Delta W_{\lambda} $$
$$y^{(n+1)}= (I + \Delta \lambda [\Sigma_1 ^{-1} + \lambda \Sigma_2 ^{-1}]^{-1} \Sigma_2^{-1}) ^{-1} (y^{(n)} + L(x) \Delta W_{\lambda} )$$
$$x^{(n+1)}= (I + \Delta \lambda [\Sigma_1 ^{-1} + \lambda \Sigma_2 ^{-1}]^{-1} \Sigma_2^{-1}) ^{-1} (x^{(n)} - m + L(x) \Delta W_{\lambda} ) +m$$
\\
\\

\subsection{Results for yield and real return model}
By involving function of movement $f(x, \lambda)$, accuracy for predicting of real return has been highly increased. The validity of using particle flow methods has been proved. The trends for predicted yield are highly similar to the real trend. And prediction for yield has great performance at the years with large fluctuation but cannot mimic the value with lower fluctuation. That is because we set relative larger covariance for likelihood matrix, which means it cannot do better when the real covariance become lower. Then the corresponding cons for Particle Flow Filter is clear to see that constant likelihood function $h(x)$ is hard to satisfy the change for each points.
\begin{figure}[h!]
	\centering
	\begin{subfigure}[b]{\linewidth}
		\includegraphics[width=\linewidth]{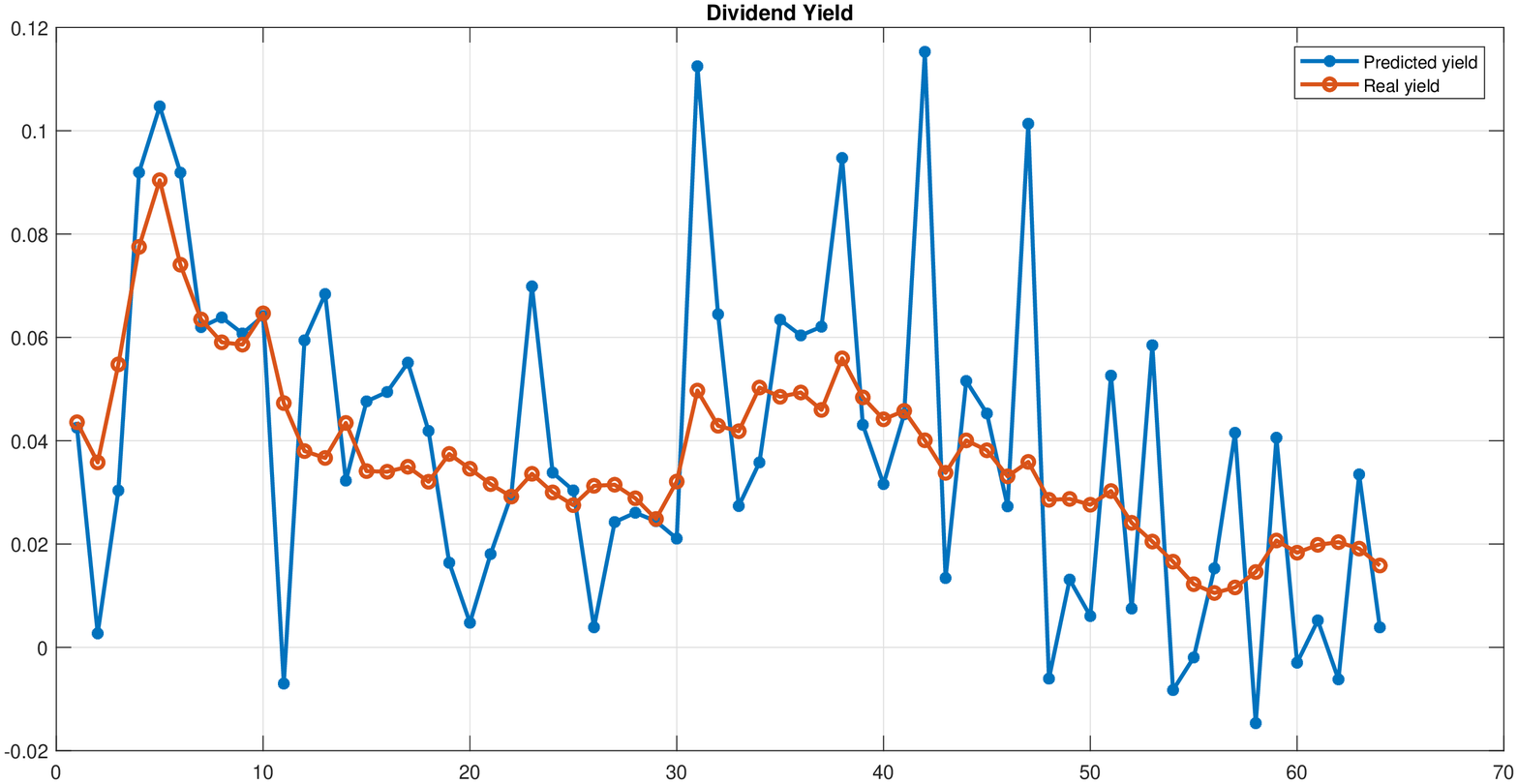}
		\caption{Yield}
	\end{subfigure}
	\begin{subfigure}[b]{\linewidth}
		\includegraphics[width=\linewidth]{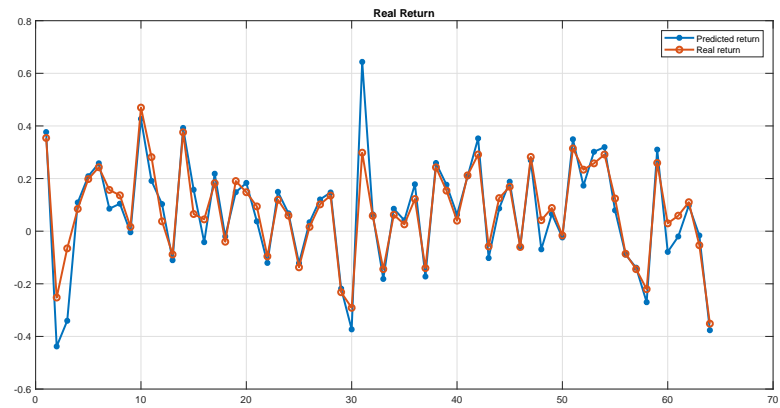}
		\caption{Return}
	\end{subfigure}
	\caption{Particle Flow Filter Results for Yield and Real Return}
\end{figure}

\end{document}